\documentclass[prd,amsmath,amssymb,nofootinbib,preprintnumbers]{revtex4-2}

\usepackage{graphicx}
\usepackage{xcolor}
\usepackage{amsthm}
\usepackage[colorlinks=true,allcolors=blue]{hyperref}
\usepackage[capitalise,nameinlink]{cleveref}

\crefname{equation}{Eq.}{Eqs.}
\Crefname{equation}{Equation}{Equations}
\creflabelformat{equation}{#2(#1)#3}
\crefname{section}{Sec.}{Secs.}
\Crefname{section}{Section}{Sections}
\crefname{subsection}{Sec.}{Secs.}
\Crefname{subsection}{Section}{Sections}
\crefname{appendix}{App.}{Apps.}
\crefname{figure}{Fig.}{Figs.}
\Crefname{figure}{Figure}{Figures}
\crefname{table}{Table}{Tables}
\crefname{benchmark}{Benchmark}{Benchmarks}

\theoremstyle{definition}
\newtheorem{benchmark}{Benchmark}

\usepackage{booktabs}
\usepackage{multirow}

\usepackage{xspace}

\newcommand{\conv}{\otimes}
\newcommand{\cauchy}[1]{\ensuremath{\mathcal{C}_{#1}}\xspace}
\newcommand{\hilbert}[1]{\ensuremath{\mathcal{H}_{#1}}\xspace}
\renewcommand{\Re}{\operatorname*{Re}}
\renewcommand{\Im}{\operatorname*{Im}}

\newcommand{\xd}{\ensuremath{x_\text{d}}\xspace}
\newcommand{\Pdet}{\ensuremath{P_\text{d}(\xd)}\xspace}
\newcommand{\yd}{\ensuremath{y_\text{d}}\xspace}
\newcommand{\xt}{\ensuremath{x_\text{t}}\xspace}
\newcommand{\yt}{\ensuremath{y_\text{t}}\xspace}
\newcommand{\zt}{\ensuremath{z_\text{t}}\xspace}
\NewDocumentCommand{\Ptruth}{s}{\ensuremath{P_\text{t}\IfBooleanF{#1}{(\xt)}}\xspace}
\NewDocumentCommand{\hatPtruth}{s}{\ensuremath{\hat{P}_\text{t}\IfBooleanF{#1}{(\zt)}}\xspace}
\NewDocumentCommand{\Psing}{s m}{\ensuremath{P_{\text{t},#2}^\text{sing}\IfBooleanF{#1}{(\xt)}}\xspace}
\NewDocumentCommand{\Prem}{s}{\ensuremath{P_\text{t}^\text{rem}\IfBooleanF{#1}{(\xt)}}\xspace}
\newcommand{\resR}[1][]{\ensuremath{\mathcal{R}\if\relax\detokenize{#1}\relax\else_{\text{#1}}\fi}\xspace}
\newcommand{\EOS}{\texttt{EOS}\xspace}

\newcommand{\amp}[1][]{\ensuremath{\mathcal{A}\if\relax\detokenize{#1}\relax\else_{\text{#1}}\fi}}

\newcommand{\GeV}{\ensuremath{\,\mathrm{GeV}}\xspace}
\newcommand{\MeV}{\ensuremath{\,\mathrm{MeV}}\xspace}
\newcommand{\keV}{\ensuremath{\,\mathrm{keV}}\xspace}

\begin{document}

\title{Unbinned Likelihoods for Probability Densities with Poles}

\author{M\'eril Reboud}
\affiliation{%
IJCLab, Pôle Théorie, CNRS/IN2P3 et Université Paris-Saclay,
F-91405 Orsay, France
}
\author{Danny van Dyk}
\affiliation{%
Institute for Particle Physics Phenomenology and Department of Physics,
Durham University, Durham DH1 3LE, UK
}
\author{Matteo Verzeletti}
\affiliation{%
Helmholtz-Institut f\"ur Strahlen- und Kernphysik (Theorie), and Cluster of Excellence ``Color meets Flavor'',
Universit\"at Bonn, D-53115 Bonn, Germany
}

\date{\today}

\preprint{EOS-2026-05, IPPP/26/79}

\begin{abstract}
Unbinned likelihood fits to detector-level data become prohibitively expensive when narrow resonances
interfere with a nonresonant amplitude. The convolution of the truth-level probability density with
the detector resolution must then resolve the resonances, whose widths are much smaller than the
resolution.
We propose a simple approach that subtracts one simple-pole term per resonance from the probability density.
This term captures the resonant contribution in closed form, and the smooth remainder is convolved numerically on a
grid whose spacing is set by the resolution alone.
The approach applies to multidimensional densities and to densities known only on the real axis.
Using two benchmark problems, we quantify the accuracy, computational cost, and statistical impact
of the approach.
We emphasise that the subtraction is a general approach to the computational problems of unbinned
likelihoods in the presence of poles in their probability density and that it is applicable to a wide variety
of problems in experimental particle physics.
\end{abstract}

\maketitle

\section{Introduction}
\label{sec:introduction}

Analyses in high-energy physics commonly infer model parameters from detector-level quantities, which are
affected by the detection efficiency and the finite detector resolution.
The likelihood of the measured data then involves the detector-level probability density function (PDF)
\Pdet, which arises from the truth-level PDF \Ptruth, including the efficiency, through the convolution
with the detector resolution \resR,
\begin{equation}
    \Pdet = \int \mathrm{d}\xt\, \Ptruth\, \resR(\xd - \xt) \equiv [\Ptruth* \conv \resR](\xd)\,,
\end{equation}
where, here and throughout this work, we assume that the resolution depends only on the offset $\xd - \xt$.
If \Ptruth is smooth, i.e., free of features that are narrow compared with the typical width
$\sigma$ of the resolution, this convolution poses no particular difficulty. It can be carried out
numerically on a grid whose spacing is a fraction of $\sigma$, e.g., by means of the discrete Fourier
transform (DFT), at a cost small enough to repeat it at every parameter point explored by a fit.

This article addresses problems in which \Ptruth is not smooth on scales below the detector resolution.
The most common source of such features is the presence of narrow resonances, whose half-widths $\gamma \ll \sigma$
would require a grid spacing of order $\gamma$ rather than $\sigma$, and whose interference with a
nonresonant amplitude precludes a closed-form convolution of the full PDF. We formalise a simple idea
that overcomes this problem: the contributions of the resonance poles are subtracted from \Ptruth, convolved with the
resolution in closed form, and added back, such that only a smooth remainder needs to be convolved
numerically on a grid set by $\sigma$ alone.

For an isolated Breit--Wigner line shape and a Gaussian resolution, the convolution is the Voigt profile~\cite{Voigt1912},
available in closed form through the Faddeeva function; Ref.~\cite{Kycia:2017gjn} derives the analogous
result for a relativistic Breit--Wigner line shape.
Mass-dependent widths or interference with a nonresonant amplitude require a numerical
convolution~\cite{CDF:2011zbc,CDF:2011ac,Wick:2011bga}. Applying the resolution to each pole of the
amplitude, as in Ref.~\cite{LHCb:2024blb}, is not exact, since the resolution acts on the density rather
than on the amplitude.

Our approach combines the closed-form and the numerical convolution. For each pole of the amplitude, we
subtract from \Ptruth a single simple-pole term, which collects the squared pole term and the singular
part of its interference with the remaining amplitude. This term is convolved with the resolution in
closed form for a Gaussian resolution, and to good approximation for the Crystal Ball
function~\cite{Oreglia:1980cs,Gaiser:1982yw,Skwarnicki:1986xj} commonly used to model detector resolutions.
The remainder is free of narrow structures and is convolved numerically on a grid whose spacing is set by
the resolution alone. Subtracting a singularity whose integral is known and integrating the smooth
remainder numerically is a classic technique of numerical quadrature~\cite{DavisRabinowitz1984}; it also
underlies the subtraction methods for infrared singularities in perturbative
calculations~\cite{Catani:1996vz,Frixione:1995ms}.
The approach applies to multidimensional PDFs and to PDFs that are known only on the real axis.

\begin{figure}[t]
    \centering
    \includegraphics{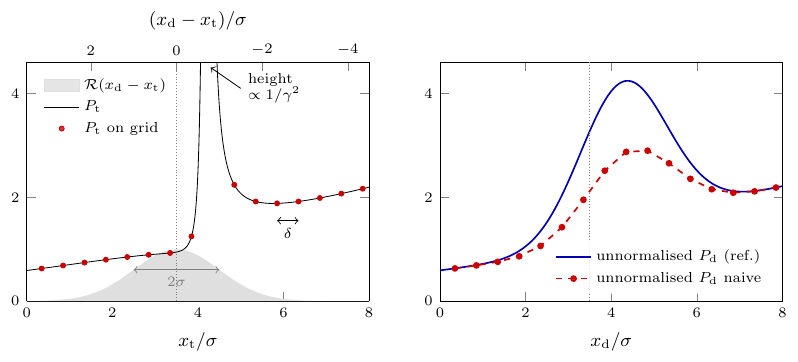}
    \caption{Illustration of the problem for a truth-level PDF with one pole of half-width
    $\gamma = \sigma/20$, which interferes with a smooth nonresonant amplitude, and a Gaussian resolution
    \resR of width $\sigma$.
    Left: \Ptruth as a function of \xt (bottom axis) and its values on a grid of spacing $\delta = \sigma/2$,
    together with the resolution kernel, in arbitrary normalisation, for the detector-level point
    $\xd = 3.5\,\sigma$ (dotted line) as a function of the offset $\xd - \xt$ (top axis).
    Right: the detector-level PDF obtained from the convolution on the grid of spacing $\delta$ (naive) and
    on a grid of spacing $\delta' = \gamma/10$ (reference).}
    \label{fig:introduction:illustration}
\end{figure}

\Cref{fig:introduction:illustration} illustrates the problem for a truth-level PDF with a single pole of
half-width $\gamma = \sigma/20$, which interferes with a smooth nonresonant amplitude, and a Gaussian
resolution. On a grid with spacing $\delta = \sigma/2$, the grid points miss the peak of \Ptruth, and the
naive convolution underestimates the height of the detector-level peak by about $30\%$ compared with a
reference obtained on a grid with spacing $\gamma/10$.
While the problem appears to be confined to within a few $\sigma$ of the pole position, this is not the case.
In an unbinned likelihood fit, \Pdet must be normalised by its integral over the fit range. A misestimate of
the area of the peak, as in \cref{fig:introduction:illustration}, therefore changes the normalised \Pdet
everywhere in the fit range, including far from the pole. This holds even if the fit range excludes
windows around the poles, since the tails of the resolution carry the peaks into the fit range; for
resolutions with power-law tails, such as the Crystal Ball function, they reach far beyond a few $\sigma$.
The misestimated normalisation therefore has the potential to bias the inference of the physics
parameters, as we demonstrate in this work.
The problem compounds in cases of multidimensional PDFs or truth-level PDFs that are computationally expensive to evaluate.

A prominent example of this problem is the rare decay $B^+\to K^+\mu^+\mu^-$, for which the LHCb experiment has recently
published the individual signal candidates together with the background, efficiency, and resolution
models~\cite{LHCb:2026suh}, enabling unbinned analyses of these data outside the collaboration.
The narrow charmonium resonances $J/\psi$ and $\psi(2S)$ enter as poles of the nonlocal amplitude and
interfere with the remaining
contributions~\cite{Deshpande:1988bd,Lim:1988yu,Ali:1991is,Kruger:1996cv,Khodjamirian:2010vf,Lyon:2014hpa,Bobeth:2017vxj,Gubernari:2020eft,Gubernari:2022hxn,LHCb:2016due}.
The $J/\psi$ has a half-width $\gamma \simeq 46\keV$, while the resolution in the dimuon mass ranges from
$2$ to $7\MeV$~\cite{LHCb:2026suh}. Resolving the $J/\psi$ over the full dimuon mass range requires a grid
of about $10^5$ points, compared with the $10^3$ to $10^4$ points that the resolution alone requires.
This physics problem inspires the benchmarks we use to assess our approach. We emphasise,
however, that our approach is not specific to $B^+\to K^+\mu^+\mu^-$: it applies to any analysis in which
narrow resonances that interfere with a nonresonant amplitude are smeared by a much broader detector
resolution.

The structure of this article is as follows. In \cref{sec:framework}, we introduce our notation, derive
the subtraction of the singular terms and the convolution of the subtracted terms, and discuss the
complications that arise in practical applications. In \cref{sec:implementation}, we describe our
implementation and present the results for two benchmark problems. We conclude in \cref{sec:conclusion}.
Our benchmarks are collected in \cref{app:benchmarks}: \cref{benchmark:B1} tests the accuracy of the
detector-level PDF against an exact reference, and \cref{benchmark:B2} serves to validate an unbinned
likelihood fit. \Cref{app:crystal-ball} discusses the Cauchy transform of the Crystal Ball function. We provide a reference implementation as part of the \EOS
software~\cite{EOSAuthors:2021xpv,EOSAuthors:2026rhs}.

\section{Framework}
\label{sec:framework}

\subsection{Notation}
\label{sec:framework:notation}

We assume that the truth-level PDF takes the form
\begin{equation}
\label{eq:framework:notation:PDF-decomp}
    \Ptruth = \frac{1}{\mathcal{N}} \sum_\lambda |\amp_\lambda(\xt)|^2\, \Phi(\xt)\,,
    \qquad
    \mathcal{N} \equiv \int \mathrm{d}\xt\, \sum_\lambda |\amp_\lambda(\xt)|^2\, \Phi(\xt)\,.
\end{equation}
Since the convolution is linear and the detector-level PDF is normalised after it,
$\mathcal{N}$ only passes through; one can set $\mathcal{N} = 1$ in the following for practical applications.
Here, the $\amp_\lambda$ represent amplitudes that we assume (for the moment) do not interfere with each other, once all other
kinematic variables are integrated over, e.g., amplitudes of different helicities, and $\Phi$ represents a phase-space factor.
This form arises in
the description of, for example, normalised differential distributions of particle decays and of normalised fiducial cross sections.
Since our treatment applies to each amplitude individually, we suppress the index $\lambda$ in the
following, and decompose each amplitude as
\begin{equation}
\label{eq:framework:notation:amplitude-decomp}
    \amp(\xt) = \amp[smooth](\xt) + \sum_j \frac{R_j}{\xt - z_j}\,.
\end{equation}
Here, $z_j = x_j - i \gamma_j$, with $\gamma_j > 0$, are the positions of the poles in the lower
complex half-plane, and $R_j$ are the (generally complex-valued) residues.
The purpose of our proposed framework is to deal with those poles in the complex plane for which
$\gamma_j \ll \sigma$, i.e., with poles that produce very narrow structure in \Ptruth that
cannot be resolved directly due to the detector resolution.
On the other hand, the description of broad structures due to poles with $\gamma_j \gtrsim \sigma$ will remain in
$\amp[smooth]$.

The truth-level PDF must further be multiplied by the detection efficiency $\varepsilon(\xt)$. The
efficiency must be a function of the truth-level variables; selections on reconstructed quantities only apply
after the convolution and enter through the fit range in \xd. Here, we assume that the efficiency function
is absorbed by $\Ptruth \to \varepsilon(\xt)\,\Ptruth$; we only discuss it explicitly where
its analytic properties matter. Strictly, \Ptruth is therefore a density with integral at most one rather
than a PDF; since the detector-level PDF is normalised after the convolution, we continue to refer to
\Ptruth as the truth-level PDF.

Throughout this work, we make the following assumptions. The smooth part of the amplitude, $\amp[smooth]$, the phase-space
factor $\Phi$, and the efficiency $\varepsilon$ are analytic in a neighbourhood of each pole $z_j$ that
includes the real axis around $x_j$, and vary there only on a scale $L \gg \gamma_j$.\footnote{%
    This excludes poles that are very close to or below a kinematic threshold, where $\Phi$ varies
    rapidly or vanishes.}
However, this does not mean that these functions need to be known in the complex plane.
If, instead, they are known only on the real axis, our approach works with the modifications discussed in \cref{sec:framework:complications}.
The proposed subtraction framework is exact for any set of distinct poles; the subtracted terms of the individual
poles and their convolutions are simply summed. The remainder, however, varies on the separations
$|x_j - x_k|$ of the poles. For the remainder to be resolved on a grid, these separations must be
large compared with the grid spacing; for the extraction of the coefficients from the real axis
(see \cref{sec:framework:complications}), they must be large compared with the extent of the set of
interpolation points.

\subsection{Subtraction of the Singular Terms}
\label{sec:framework:subtraction}

We subtract the singular terms of each (complex-valued) pole from \Ptruth. To this end, we introduce \hatPtruth as the
analytic continuation of \Ptruth into the complex plane,
\begin{equation}
\label{eq:framework:subtraction:Phat}
    \hatPtruth \equiv \frac{\Phi(\zt)}{\mathcal{N}}\, \amp(\zt)\, \big[\amp(\zt^*)\big]^*\,.
\end{equation}
By construction, it coincides with \Ptruth on the real axis and satisfies $\hatPtruth = [\hatPtruth*(\zt^*)]^*$. Near
$z_j$, only the factor $\amp(\zt)$ is singular; \hatPtruth therefore has a simple pole at $\zt = z_j$ and, by
reflection, one at $\zt = z_j^*$ with the conjugate residue. For each pole we define the coefficient $c_j$
in terms of the residue in the complex plane,
\begin{equation}
\label{eq:framework:subtraction:F}
    c_j \equiv 2 \operatorname*{Res}_{z = z_j} \hatPtruth*(z) = \frac{\Phi(z_j)}{\mathcal{N}} \left[ 2\, \big[\amp^{(j)}(z_j^*)\big]^*\, R_j
        + \frac{i\,|R_j|^2}{\gamma_j} \right]\,,
\end{equation}
where $\amp^{(j)}(z) \equiv \amp(z) - R_j/(z - z_j)$ is analytic in a neighbourhood of $z_j$ and varies on
the scale $L$ and on the separations $|x_j - x_k|$; in the following, we take $L$ to include the
latter. Each pole coefficient $c_j$ thus collects both the squared pole term and the singular part of its
interference with the remaining amplitude. We decompose the PDF as
\begin{equation}
\label{eq:framework:subtraction:psing}
    \Ptruth = \Prem + \sum_j \Psing{j}\,,
    \qquad
    \Psing{j} \equiv \Re\left[\frac{c_j}{\xt - z_j}\right]\,.
\end{equation}
The remainder \Prem is the restriction to the real axis of
$\hatPtruth*(z) - \sum_j \big[\tfrac{c_j/2}{z - z_j} + \tfrac{c_j^*/2}{z - z_j^*}\big]$, which is
analytic in a neighbourhood of every pole, regardless of their separations. Hence, \Prem has no
structure on the scales $\gamma_j$, and an efficient numerical convolution with \resR is possible on a
grid whose spacing is set by $\sigma$ and $L$ alone.

To obtain the full result of the convolution, we now need to deal with the singular terms.
We obtain for each pole individually the exact result
\begin{equation}
\label{eq:framework:subtraction:addback}
    [\Psing*{j} \conv \resR](\xd) = \Re\left[ c_j\, \cauchy{\resR}(\xd - z_j) \right]\,,
    \qquad
    \cauchy{\resR}(\zeta) \equiv \int \mathrm{d}u\, \frac{\resR(u)}{\zeta - u}\,,
    \quad \Im \zeta > 0\,,
\end{equation}
where we introduce the Cauchy transform\footnote{%
    Also known as the Cauchy--Stieltjes transform; it equals the Stieltjes transform up to a sign
    convention~\cite{CimaMathesonRoss2006}.
}~$\cauchy{\resR}$ that depends only on the resolution:
\begin{itemize}
    \item For a Gaussian resolution of width $\sigma$,
    \begin{equation*}
        \resR[G](u) = \frac{1}{\sqrt{2\pi\sigma^2}} e^{-u^2/(2\sigma^2)}\,,
    \end{equation*}
    we use the well-known result
    \begin{equation*}
        \cauchy{\resR[G]}(\zeta) = -i\,\sqrt{\frac{\pi}{2 \sigma^2}}\, w\left(\frac{\zeta}{\sqrt{2 \sigma^2}}\right)\,,
    \end{equation*}
    where $w(z)$ denotes the Faddeeva function $w(z) = e^{-z^2}\operatorname{erfc}(-iz)$.

    \item For a Crystal Ball resolution, as defined in \cref{eq:app:cb:def}, the exact Cauchy transform
    requires an incomplete Faddeeva function, for which we are not aware of an implementation
    of a quality comparable to that of the complete Faddeeva function $w(z)$.
    We derive an approximate result for its Cauchy transform $\cauchy{\resR[CB]}$ in \cref{app:crystal-ball},
    by expressing the Crystal Ball resolution as a superposition of the Gaussian kernel $\resR[G]$
    and a correction term $\Delta$.

    \item For resolutions that are described as linear combinations of the above two possibilities,
    we use linearity of the Cauchy transform.
\end{itemize}
Regardless of the composition of the resolution, all dependence on the physical parameters of the singular terms
\Psing{j} enters \Pdet only through the coefficients $c_j$.

We recommend keeping track of the numerically convolved remainder $[\Prem* \conv \resR]$ on a grid and interpolating it to the points at which the detector-level PDF is needed, e.g., the observed events.
The Cauchy transforms $\cauchy{\resR}(\xd - z_j)$, which carry the peaks, are instead evaluated in closed form at
these points. The grid spacing therefore needs to resolve only the scales $\sigma$ and
$L$ on which the convolved remainder varies, but not the shape of the smeared peaks.\\

For PDFs that depend on other kinematic variables beyond \xt, which we collectively denote by \yt, the
residues $R_j$ become functions of \yt. We require the pole positions $z_j$ to be independent of \yt.
The coefficients $c_j(\yt)$ then depend on \yt, and the singular terms read
$\Re[c_j(\yt)/(\xt - z_j)]$. Interference terms $\Re[\amp_\lambda \amp_{\lambda'}^*]$, weighted by functions of
\yt, contribute to $c_j(\yt)$ in the same way, whether one or both of the interfering amplitudes carry the
pole $z_j$. The resolution may act on \yt as well.
Our approach easily generalises to a resolution of the form
$\resR(u, v) = \sum_\nu h_\nu(v)\, \resR_\nu(u)$, where the $\resR_\nu$ are kernels in the offset $u$ of the pole
variable and the $h_\nu$ are kernels in the offsets $v$ of the other variables or, for a resolution
that does not act on them, simply $h_\nu(v) = \delta(v)$.
In this case, the singular terms are convolved as
\begin{equation}
\label{eq:framework:subtraction:addback-multi}
    [\Psing*{j} \conv \resR](\xd, \yd) = \sum_\nu \Re\Big[ (c_j \conv h_\nu)(\yd)\, \cauchy{\resR_\nu}(\xd - z_j) \Big]\,,
\end{equation}
where $c_j \conv h_\nu$ denotes the convolution in \yt.
This allows the use of efficiencies $\varepsilon(\xt, \yt)$ that do not factorise in \xt and \yt, which enter $c_j(\yt)$ pointwise in
\yt.
For a factorised resolution, $\resR(u, v) = \resR_x(u)\,\resR_y(v)$, the right-hand side of \cref{eq:framework:subtraction:addback-multi}
reduces to
\begin{equation}
\Re[(c_j \conv \resR_y)(\yd)\, \cauchy{\resR_x}(\xd - z_j)]\,.
\end{equation}
The convolved coefficients $(c_j \conv h_\nu)$, which vary only on the scales of the other variables,
are stored on the grid alongside the remainder and interpolated in the same way, while the Cauchy
transforms $\cauchy{\resR_\nu}(\xd - z_j)$ are evaluated in closed form.

\subsection{Complications}
\label{sec:framework:complications}

We identify six complications to the immediate application of our idea, and describe how we address
them.

\noindent
\textbf{1) Evaluation in the complex plane.}~%
The computation of the coefficient $c_j$ requires the evaluation of its building blocks
$\amp^{(j)}$, $\Phi$, and $\varepsilon$ at complex arguments.
If their evaluation in the complex plane is possible, then the evaluation of $c_j$ follows directly from
\cref{eq:framework:subtraction:F}.
An efficiency $\varepsilon$ parametrised analytically, e.g., as a polynomial in the truth-level variables, can be
evaluated at complex arguments directly. The evaluation in the complex plane is not possible, however,
if the nonresonant amplitude $\amp^{(j)}$ or the efficiency $\varepsilon$ is known only on the real axis, e.g., as a table.
In that case, $c_j$ can nevertheless be extracted from the values of \Ptruth on the real axis alone.
To this end, we introduce the function
\begin{equation}
\label{eq:framework:complications:g}
    g_j(\xt) \equiv |\xt - z_j|^2\, \Ptruth
        = \frac{\Phi(\xt)\,\varepsilon(\xt)}{\mathcal{N}}\, \big|(\xt - z_j)\, \amp^{(j)}(\xt) + R_j\big|^2\,.
\end{equation}
It is free of the pole and varies only on the scale $L$. Its analytic continuation $\hat{g}_j$ yields
the coefficient exactly,
\begin{equation}
\label{eq:framework:complications:c-from-g}
    c_j = \frac{i}{\gamma_j} \left[\hat{g}_j(z_j^*)\right]^*\,,
    \qquad
    \hat{g}_j(z) = (z - z_j)(z - z_j^*)\,\hatPtruth*(z)\,,
\end{equation}
since $\hat{g}_j(z_j^*) = 2i\gamma_j\,\big[\operatorname*{Res}_{z = z_j}\hatPtruth*\big]^* = i\gamma_j\, c_j^*$. For practical applications, we recommend interpolating $g_j$ through the $K$ grid points closest to $x_j$ by the interpolating polynomial of degree $K - 1$, and evaluating it at $z_j^* = x_j + i\gamma_j$.
Since \Ptruth is sampled on the grid anyway, this requires no additional evaluations. For a grid
spacing $\delta \ll L$, the error of the interpolation decreases like $(\delta/L)^{K}$. The interpolation nodes must not
enclose an edge of the phase space or another pole.
The same procedure applies to each amplitude and each interference term individually, and pointwise in \yt.

\noindent
\textbf{2) Pole in $\boldsymbol{\xt^2}$, resolution in $\boldsymbol{\xt}$.}~%
In some applications, the resonances are naturally described as poles in
$\xt^2$, with $\amp \ni R_j/(\xt^2 - s_j)$ and $s_j = m_j^2 - i m_j \Gamma_j$ with $m_j, \Gamma_j > 0$,
while the resolution acts as a convolution in \xt; this necessitates working in the latter variable. The change of
variables multiplies the PDF by the Jacobian $2\xt$, which is smooth in the
vicinity of the pole and can be absorbed into $\Phi$. Partial fractioning in \xt yields
\begin{equation}
\label{eq:framework:complications:partial-fractions}
    \frac{R_j}{\xt^2 - s_j} = \frac{R_j}{2 z_j}\left[\frac{1}{\xt - z_j} - \frac{1}{\xt + z_j}\right]\,,
    \qquad
    z_j \equiv \sqrt{s_j}\,,\quad \Re z_j > 0\,.
\end{equation}
The first term is a pole of the form used in \cref{sec:framework:notation} at $\xt = z_j$
with residue $R_j/(2z_j)$ and half-width $\gamma_j = -\Im z_j = m_j \Gamma_j/(2 \Re z_j) \simeq \Gamma_j/2$.
The mirror pole at $\xt = -z_j$ lies at a distance of about $m_j$ from the lower edge of the phase
space and is absorbed into $\amp[smooth]$.
The pole position should be taken as the exact square root $z_j$;
approximating it as $m_j - i\Gamma_j/2$ shifts the pole by $\mathcal{O}(\Gamma_j^2/m_j)$ and leaves an
unsubtracted remainder.

\noindent
\textbf{3) Circular convolution and padding.}~%
Since the DFT implements a circular convolution, the grid must
extend beyond the physical range by more than the reach of \resR; otherwise, contributions from one
end of the grid wrap around to the other.
In addition, the kernel is sampled only on offsets up to half the extent of
the grid, and is cut off beyond. The subtraction adds a subtlety:
the singular terms \Psing{j} are not truncated to the support of \Ptruth, since their convolution
would otherwise no longer be known in closed form. Consequently, \Prem does not vanish in the
padding, where it equals $-\sum_j \Psing{j}$. This is permissible, since the convolution of \Prem is
correct at all points farther than the reach of the resolution from the ends of the grid; the
physical region must lie entirely within these points. For resolutions with power-law tails, this
reach is not sharply defined; the truncated tails leave an error that decreases with the padding
(see the discussion in \cref{sec:implementation:B1}). Moreover, the subtraction removes the
effect of the cut-off of the kernel on the contributions of the poles, since the singular terms are
convolved through the Cauchy transform of the full, untruncated kernel. For resolutions with heavy
power-law tails, such as the Crystal Ball function, the cut-off of the kernel is visible in the naive
convolution but not in the subtracted one (as illustrated in \cref{sec:implementation:B1}).

\noindent
\textbf{4) Normalisation.}~%
If an application requires knowledge of the \emph{truth-level} normalisation $\mathcal{N}$, its computation involves an integral over the narrow peaks.
We propose evaluating this integral with the same subtraction: the integrals of the singular terms are known in closed form in
terms of complex logarithms, while the remainder is free of narrow structures and is integrated numerically.
In a fit, moreover, the \emph{detector-level} PDF must be normalised by its own integral over the fit range rather
than by $\mathcal{N}$, such that one accounts for the probability that the resolution moves into or out of the fit range.
This applies equally to fits restricted to windows, or more generally to unions of non-overlapping boxes in the observables, which include only the events within them.
A convolution that misrepresents the peaks also misrepresents their share of this integral;
this modifies the detector-level PDF everywhere in the fit range, including far from the poles.
The problem persists even if windows around the resonances are excluded from the fit, as long as the peaks reach into the fit range
through the tails of the resolution.
The grid must therefore cover all truth-level support whose resolution
tails reach the fit range, including the poles. With the subtraction, the integral is the sum of the
integral of the interpolated convolved remainder and that of the convolved singular terms.

\noindent
\textbf{5) Piecewise resolution.}~%
The resolution may depend on the position $\xt$ rather than the offset $\xd - \xt$.
If it is defined piecewise in the detector-level variable, then the detector-level PDF within a region $r$ is $[\Ptruth* \conv \resR_r](\xd)$ for
\xd in that region. Our approach then applies to each region separately: \Ptruth is evaluated on the
grid once, its remainder is convolved with each kernel $\resR_r$, and the singular terms through the
Cauchy transform of each kernel. Each kernel acts on the full truth-level support, such that all poles
contribute to each region through the tails of the resolution. The integral over the fit range is the
sum of the integrals over its intersections with the regions. A resolution that depends continuously
on the truth-level position is beyond the scope of this work.

\noindent
\textbf{6) Edges and thresholds.}~%
The subtraction treats poles only.
Structures on scales smaller than $\sigma$ that remain in \Prem include thresholds and edges of the phase space (where \Ptruth jumps or $\Phi$ vanishes).
They are convolved on the grid, and the error of the detector-level PDF grows close to them.
For resolutions with power-law tails, they also leave small kinks far from them, where the kernel is cut off
(as illustrated in \cref{sec:implementation:B1}). The interpolation nodes of the coefficients must not enclose them.
Poles close to a threshold are excluded by the assumptions made in \cref{sec:framework:notation}.

\section{Implementation and Benchmarking}
\label{sec:implementation}

We have implemented our approach within the \EOS software~\cite{EOSAuthors:2021xpv,EOSAuthors:2026rhs}.
The implementation is available since \EOS version 1.1.1~\cite{EOS:v1.1.1}.
For the efficient and accurate evaluation of the Faddeeva function, we have partially incorporated the code
of the \texttt{libcerf} library~\cite{libcerf} within \EOS, after translating it to C++.
Besides \texttt{libcerf}, we make use of \texttt{NumPy}~\cite{Harris:2020xlr}, \texttt{SciPy}~\cite{Virtanen:2019joe},
\texttt{mpmath}~\cite{mpmath}, \texttt{matplotlib}~\cite{Hunter:2007ouj}, and \texttt{dynesty}~\cite{Speagle:2019ivv} in the course of this analysis.

We benchmark our implementation using the two problems introduced in \cref{app:benchmarks}. Both problems
are inspired by the physics application that we have in mind to solve: the interpretation of the signal candidates
for the decay $B^+\to K^+\mu^+\mu^-$ provided by the LHCb experiment~\cite{LHCb:2026suh}. They are accessible
as built-in signal PDFs that are registered on demand only\footnote{%
    By calling \texttt{eos.register\_benchmarks()} within Python, or by setting the environment variable
    \texttt{EOS\_REGISTER\_BENCHMARKS}.
} through the names
\begin{center}
    \texttt{benchmark::B1@RvDV:2026A}\qquad and\qquad\texttt{benchmark::B2@RvDV:2026A}\,.
\end{center}
The former corresponds to a constant contribution, interfering with two narrow resonances resembling the $J/\psi$ and $\psi(2S)$;
this problem is meant to illustrate the accuracy of our approach, since the simple interference patterns provide
a controllable reference result.
The latter corresponds to a smooth, slowly varying component dependent on three real-valued physics parameters, interfering with the
two narrow resonances;
this problem is meant to study the performance of the unbinned likelihood, in terms of accuracy and computational costs,
in a physics-inspired scenario.
This scenario resembles an application of our approach for creating an unbinned likelihood from the recently published 
signal candidates, and efficiencies \& resolution models by the LHCb experiment~\cite{LHCb:2026suh}.
The analysis file and ancillary scripts used to obtain the results shown below are provided as part of the
supplementary material~\cite{EOS-DATA-2026-05}.

\subsection{Benchmark B1}
\label{sec:implementation:B1}

\begin{figure}[t]
    \centering
    \includegraphics[width=.8\textwidth]{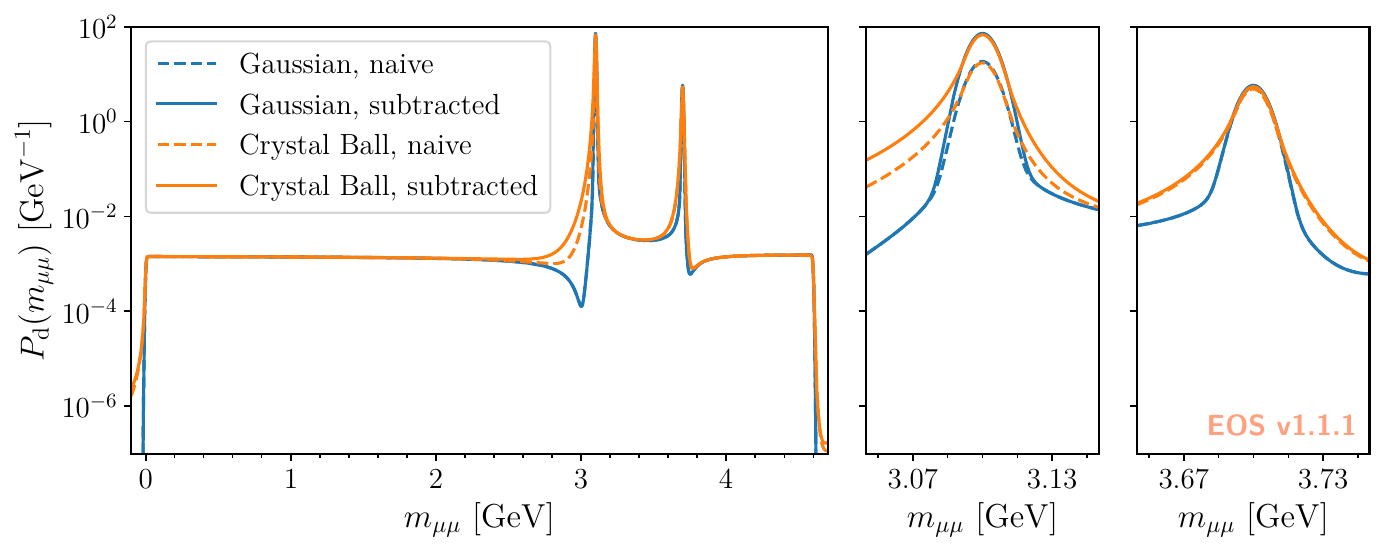}
    \caption{Detector-level PDFs of \cref{benchmark:B1} for a Gaussian (blue) and a Crystal Ball (orange)
    resolution, obtained with the naive numerical convolution (dashed) and with the subtraction of the
    singular terms (solid). The right panels show the regions within $\pm 10\sigma$ of the
    $J/\psi$-like and $\psi(2S)$-like poles; they share the vertical axis of the left panel.
    The subtracted results are visually indistinguishable from our reference results; we
    illustrate the relative deviation in \cref{fig:implementation:B1-relative-error}.
    }
    \label{fig:implementation:B1-PDF}
\end{figure}

\noindent
\textbf{Grid and reference.}~%
We evaluate the detector-level PDFs of \cref{benchmark:B1} on a grid of
$N = 8192$ points with a default spacing of $\delta = 0.62\MeV \simeq \sigma / 8$,
where $\sigma$ is the resolution scale.
The grid covers the range $\xd \in [-0.25, 4.849]\GeV$, corresponding to the full phase space
with a padding of about $50\sigma$ on either side.
Our implementation obtains the coefficients $c_j$ from the grid with $K = 7$ interpolation nodes,
unless they are provided in closed form. For \cref{benchmark:B1}, the coefficients obtained in this way
agree with their closed form to a relative precision of $10^{-13}$ for $K = 5$ and
$3\cdot 10^{-16}$ for $K = 7$, i.e., essentially at machine precision.
For both resolutions, we compare the naive numerical convolution of \Ptruth with the
convolution after the subtraction of the singular terms.

The reference is the exact detector-level PDF, which we obtain without a grid. For \cref{benchmark:B1}, \Ptruth
equals $\mathcal{N}^{-1} + \sum_j \Re[c_j/(\xt - z_j)]$ inside the kinematically allowed region $[0, W]$ and vanishes outside. We convolve the pole
terms through the Cauchy transform of the resolution, which we evaluate with the Faddeeva function of
\texttt{SciPy} for the Gaussian and by adaptive numerical integration for the Crystal Ball. The
remaining terms, $\mathcal{N}^{-1}$ inside $[0, W]$ and $-\sum_j \Re[c_j/(\xt - z_j)]$ outside, are free
of poles, and we convolve them by adaptive numerical integration. For the Gaussian, the reference
thus shares the coefficients $c_j$ and the closed form of $\cauchy{\resR[G]}$ with the subtracted
convolution, but neither the implementation of the Faddeeva function nor the treatment of the
remainder; the comparison therefore tests the grid-based part of our approach and its
implementation. For the Crystal Ball, the reference is independent of our approach.
We have cross-checked the Gaussian reference against a direct adaptive quadrature of \Ptruth
convolved with $\resR[G]$ in 30-digit arithmetic with \texttt{mpmath}~\cite{mpmath}, which uses
neither the coefficients $c_j$ nor the Faddeeva function; both agree to $10^{-15}$.

\begin{table}[b]
\centering
\begin{tabular}{lcc}
    \toprule
    padding                             & $\xd \in [0.2, 2.8]\GeV$      & $\xd \in [4.0, 4.4]\GeV$      \\
    \midrule
    $0.25\GeV$                          & $(1$--$2.5)\cdot 10^{-6}$     & $2.8\cdot 10^{-5}$            \\
    $1.25\GeV$                          & $1.2\cdot 10^{-6}$            & $(2$--$3)\cdot 10^{-6}$       \\
    $4.25\GeV$                          & $3.5\cdot 10^{-7}$            & $(0.3$--$1.2)\cdot 10^{-6}$   \\
    \bottomrule
\end{tabular}
\caption{Floor of the relative deviation of the subtracted detector-level PDF of \cref{benchmark:B1} from the
exact result far from the poles, for the Crystal Ball resolution, as a function of the padding of the grid at
the fixed spacing $\delta = 0.62\MeV$.}
\label{tab:implementation:B1-floor}
\end{table}

\begin{figure}[t]
    \centering
    \includegraphics[width=.8\textwidth]{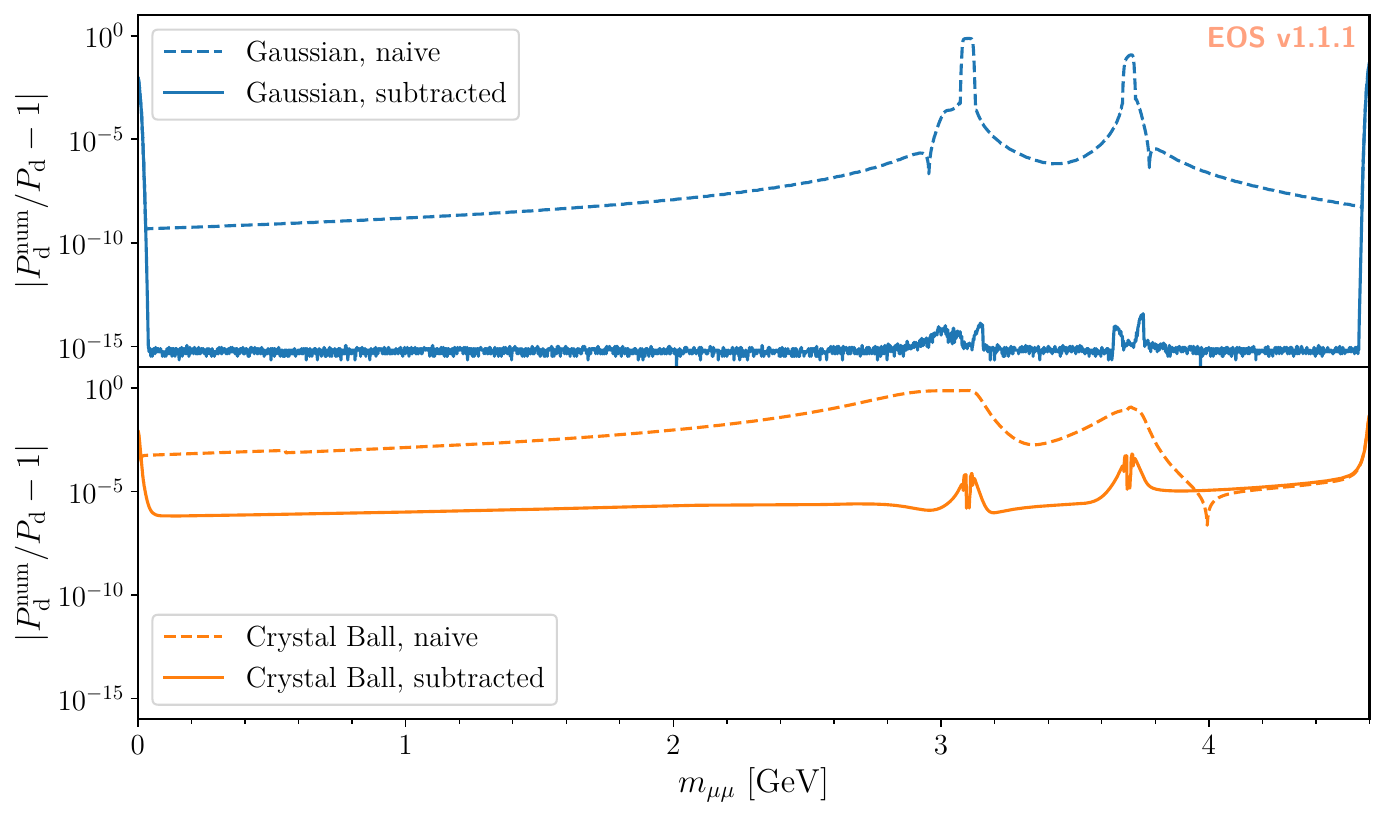}
    \caption{Relative deviation of the numerical detector-level PDFs of \cref{benchmark:B1} from the exact
    result, for a Gaussian (top) and a Crystal Ball (bottom) resolution, obtained with the naive
    numerical convolution (dashed) and with the subtraction of the singular terms (solid), where
    $P_\text{d}^\text{num}$ denotes the numerical detector-level PDF.}
    \label{fig:implementation:B1-relative-error}
\end{figure}

\noindent
\textbf{Accuracy.}~%
Both detector-level PDFs are normalised with the exact normalisation of \Ptruth, which isolates the
local error of the convolution. We show them in \cref{fig:implementation:B1-PDF}, and their relative
deviations from the reference result in \cref{fig:implementation:B1-relative-error}. Since the naive
convolution does not resolve the widths $\gamma_j$, it deviates from the reference result within
$\pm 1\sigma$ of the poles by up to $74\%$ (around the $J/\psi$-like pole) and $11\%$ (around the
$\psi(2S)$-like pole), for both resolutions. Away from the poles, it agrees well with the reference result
for the Gaussian resolution. For the Crystal Ball resolution, it underestimates the exact result by up
to $6\%$ even far below the poles, for $\xd \in [0.1, 2.5]\GeV$, since the unresolved peaks are
underrepresented in the power-law tails of the resolution.

With the subtraction, the deviations for the Gaussian resolution are at the level of the machine precision:
they stay below $2\cdot 10^{-15}$ within $\pm 1\sigma$ of the poles, and below $4\cdot 10^{-14}$ for
$\xd \in [0.05, W - 0.05]\GeV$, i.e., everywhere except close to the edges of the phase space. This is
expected for \cref{benchmark:B1}: the remainder is constant inside $[0, W]$ and, away from the edges, is convolved on the grid without error.
The singular terms are evaluated in closed form at each point. Since the reference shares them, the
coefficients $c_j$ and the closed form of $\cauchy{\resR[G]}$ are tested only by the cross-check with
\texttt{mpmath}; a remainder that varies inside the physical range is tested with \cref{benchmark:B2}.

For the Crystal Ball resolution, the subtracted result has two separate sources of error. Within about
$\pm 10\sigma$ of the poles, the error stems from truncating the expansion of the Cauchy transform of the
Crystal Ball function at $\mathcal{O}(\gamma)$; see \cref{app:crystal-ball}. It is largest at
$\xd - x_j \approx +3\sigma$, where it amounts to $7.4\cdot 10^{-5}$ for the
$J/\psi$-like and $6.4\cdot 10^{-4}$ for the $\psi(2S)$-like pole, and it does not exceed
$4\cdot 10^{-6}$ and $4\cdot 10^{-5}$, respectively, within $\pm 1\sigma$.
The ratio between the two poles, $8.7$, agrees with the expected scaling
$(\gamma_2/\gamma_1)^2 = 9$, and the error is unchanged when the grid spacing is halved\footnote{%
    On alternative grids with spacings of $\sigma/2$ and $\sigma/4$, the subtracted results near the poles and far from
    the edges agree with those discussed here to the precision quoted; the errors that stem from the edges of the
    phase space, where the step of \Ptruth is sampled on the grid, grow to at most
    $1.3\cdot 10^{-4}$. The finer spacing is therefore required by the edges only, not by
    the subtraction.
} or the padding is enlarged.
Near the poles, $c_j$ is dominated by $i|R_j|^2/\gamma_j$, such that the detector-level PDF probes the
imaginary part of the truncation error, $\tfrac{\gamma_j^2}{2}\pi\Delta''(y)$ with $y = \xd - x_j$, relative to
$\pi\resR[CB](y)$; away from the junctions, this ratio peaks at $y \approx +3\sigma$. The larger error of $\cauchy{\resR[CB]}$ at the
junction $y = -\alpha_\text{L}\sigma$ (\cref{app:crystal-ball}) resides in its real part and enters only
through the interference term of $c_j$.

Far from the poles, the error reaches a floor that is set by the extent of the grid rather than by its
spacing: the power-law tails of the kernel ($n_\text{L} = 3$) are cut off and wrapped around by the
circular convolution. \Cref{tab:implementation:B1-floor} lists this floor for the padding of
$0.25\GeV$ used here and for larger paddings at the same spacing; halving the spacing leaves it
unchanged. With the larger paddings, the floor for $\xd \in [4.0, 4.4]\GeV$ is instead set by the
discretisation of the edge at $W$ and depends on the position of the grid nodes relative to it. The naive result shows a similar floor for $\xd \in [4.0, 4.4]\GeV$, about
$2\cdot 10^{-5}$. This floor is a property of the finite periodic grid, not of the
subtraction. Resolutions with heavy tails therefore require more padding than Gaussian ones.

In the circular convolution, the sampled kernel is cut off at half the
extent of the grid, $\pm 2.55\GeV$. For the Crystal Ball resolution, the heavy power-law
tail with $n_\text{L} = 3$ makes this cut-off visible as kinks at $m_j - 2.55\GeV$ in the naive
result. These kinks are absent after the subtraction, as anticipated in
\cref{sec:framework:complications}. Kinks at $W - 2.55\GeV$ and $2.55\GeV$, caused
by the edges of the phase space, remain at the level of the floor, since the edges are not treated by
the subtraction. Close to the edges of the phase space, the deviations of both methods grow
to the percent level.

\noindent
\textbf{Normalisation over the fit range.}~%
We normalise each detector-level PDF by its own integral, as required in a fit.
Over $[0, W]$, the integral of the naive detector-level PDF amounts to only $32\%$ of the
exact one for both resolutions. Normalised
in this way, the naive detector-level PDF is wrong by a factor of about $3$ everywhere in
$[0, W]$. The integrals of the subtracted detector-level PDFs agree with the exact one to better than
$5\cdot 10^{-7}$.
The deficit observed for the naive detector-level PDF is attributable almost entirely to the unresolved peaks.
Excluding the charmonium windows chosen as \xd in $[2.83,\, 3.32] \cup [3.54,\, 3.87]\,\GeV$, we find similar relative deviations from the reference result for both the naive and subtracted methods applied to the Gaussian resolution and for the subtracted case applied to the Crystal Ball resolution, all with a value of $5 \cdot 10^{-5}$.
This common value stems from the unresolved edges of the phase space rather than the peaks.
For the windowed cases involving a Gaussian resolution, the application of our method is therefore not essential.
However, even for the windowed case, the naive method applied to the Crystal Ball resolution shows relative deviations up to $43\%$,
largest at the lower edge of the removed $J/\psi$ window, because of the misrepresented peaks carried into the selected window by the power-law tails.
Here, the subtraction approach is needed to obtain reliable results, even in the windowed case, when using the grid size specified above.

\noindent
\textbf{Grid size and computational cost.}~%
To match the accuracy of the subtracted Crystal Ball result at the $J/\psi$-like peak, the naive approach requires a grid size
of $N \gtrsim 1.8\cdot 10^5$.
For this naive approach and over the full range, the deviation saturates at about $10^{-3}$ for $N \gtrsim 1.6\cdot 10^5$, close to $m_1 - 2.55\GeV$, where the circular convolution cuts off the kernel centred on the $J/\psi$-like peak.
A finer grid does not lower this floor, but more padding does: at the same spacing, a padding of $1.25\GeV$, i.e., $N \approx 3.6\cdot 10^5$, lowers it to $3\cdot 10^{-5}$.
This is below the largest deviation of the subtracted result, $6.4\cdot 10^{-4}$, which stems from the truncation of the expansion close to the $\psi(2S)$-like pole.

The computational cost of evaluating an unbinned likelihood involving our subtracted detector-level PDF for one fixed set of parameters
includes the cost for the evaluation of \Ptruth on the grid, the convolution, and the evaluation of the convolved \Pdet at the provided events.
We measure it on a single core of an AMD EPYC 74F3 processor. For the subtracted detector-level
PDF at $N = 8192$, the evaluation of \Ptruth on the grid takes $3$\,ms, while the fast Fourier
transforms take only $0.2$\,ms. The cost is therefore set by the number of grid points, rather than by
the $N \log N$ scaling of the fast Fourier transform. The naive detector-level PDF costs $55$ times as
much at $N \approx 1.8\cdot 10^5$, and $113$ times as much at $N \approx 3.6\cdot 10^5$ with the larger
padding. This grows faster than the number of grid points, since the evaluation of \Ptruth costs
$0.4\,\mu\mathrm{s}$ per point at $N = 8192$ but $1.0\,\mu\mathrm{s}$ per point on the large grids. At the events, the subtracted
detector-level PDF evaluates the Cauchy transforms of the resolution, which takes $0.3\,\mu\mathrm{s}$
per event, against $0.01\,\mu\mathrm{s}$ for the interpolation of the naive one. For $10^4$ events,
as within a typical $q^2$ window, the subtracted detector-level PDF thus spends $3$\,ms on the events,
comparable to the evaluation of \Ptruth on the grid, and the naive one $0.1$\,ms.
Including the events, the naive detector-level PDF at $N \approx 1.8\cdot 10^5$ costs about $30$ times as much. The tables of the Cauchy transform of the Crystal Ball function
(see \cref{app:crystal-ball}) are built once per resolution, in $0.6$\,s.
For comparison, the differential decay rate of $B^+\to K^+\mu^+\mu^-$ in \EOS, with the local form
factors parametrised following Ref.~\cite{Bharucha:2015bzk} and the nonlocal form factors of
Ref.~\cite{Gubernari:2022hxn}, takes $16\,\mu\mathrm{s}$ per point, about $40$ times as long as the
$0.4\,\mu\mathrm{s}$ of the \Ptruth of \cref{benchmark:B1}.

\subsection{Benchmark B2}
\label{sec:implementation:B2}

\noindent
\textbf{Grid and accuracy.}~%
We evaluate the detector-level PDFs of \cref{benchmark:B2} on a grid of $2048 \times 512$ points
in $(m_{\mu\mu}, \cos\theta_\ell)$. In $m_{\mu\mu}$, the grid covers $[2m_\mu - 0.25, m_B - m_K + 0.25]\GeV$
at a spacing of $2.48\MeV \approx \sigma/2$. In $\cos\theta_\ell$, it covers
$|\cos\theta_\ell| \leq 1.1995$ at a spacing of $1/213 \approx \sigma_{\cos\theta}/4$,
chosen such that the edges of the phase space at $\cos\theta_\ell = \pm 1$ lie halfway between grid
nodes. Compared with a grid of twice as many points in each variable, the subtracted detector-level
PDFs deviate by at most $2\cdot 10^{-5}$ for $|\cos\theta_\ell| < 0.9$, including within
$100\MeV$ of the poles, and by up to $5\cdot 10^{-4}$ for $|\cos\theta_\ell| \geq 0.9$,
where the smeared step at $\cos\theta_\ell = \pm 1$ is interpolated. The naive detector-level PDFs
deviate from the subtracted ones on the finer grid by up to $85\%$ within $100\MeV$ of the poles for both resolutions, and by up to
$69\%$ far from the poles for the Crystal Ball resolution.
This confirms that a grid
spacing of about $\sigma/2$ suffices, as anticipated in \cref{sec:framework:subtraction}. The coefficients $c_j(\cos\theta_\ell)$ are obtained from the grid
with $K = 7$ interpolation nodes; the resulting detector-level PDFs agree with those obtained from the
closed-form coefficients to better than $10^{-12}$.

\noindent
\textbf{Normalisation over the window.}~%
Within the $q^2$ window, the detector-level PDF is normalised by its own integral over the window;
as discussed in \cref{sec:framework:complications}.
In our implementation, we integrate the multilinear interpolation of the grid analytically, i.e., without resorting to (potentially adaptive)
quadrature rules.
With poles, we add the integral of the convolved singular terms. In $m_{\mu\mu}$, we divide the window into equal subintervals
no wider than one grid spacing and apply a four-point Gauss--Legendre rule on each; in $\cos\theta_\ell$, we integrate the interpolated coefficients analytically.
At the parameter point of \cref{tab:benchmarks:B2}, which we also use to generate the pseudo-data below, we compare the probability
that an event falls into the $q^2$ window, as given by our implementation, with an independent calculation:
$\int \mathrm{d}m\, p(m) \int_\text{window} \mathrm{d}m'\, \resR(m' - m)$, where $p$ denotes the truth-level PDF in $m_{\mu\mu}$, marginalised over $\cos\theta_\ell$.
For the subtracted detector-level PDF, both calculations agree to $2.3\cdot 10^{-6}$ for the Gaussian and to $4.3\cdot 10^{-6}$ for the
Crystal Ball resolution.
For the naive detector-level PDF they agree to $1.7\cdot 10^{-6}$ for the
Gaussian resolution, since the peaks lie outside the window.
However, for the Crystal Ball resolution the naive detector-level PDF misses $4\%$ of the
probability, since its power-law tails carry the unresolved $J/\psi$-like peak into the window.

\noindent
\textbf{Fits to large pseudo-data sets.}~%
We illustrate the impact of the approach with one large pseudo-data set per resolution,
generated at the parameter point of \cref{tab:benchmarks:B2}, of ten times the size of the LHCb
Run~1+2 sample. Within the $q^2$ window, they comprise $80\,255$ events for the Gaussian and
$84\,591$ events for the Crystal Ball resolution.
These pseudo-data sets are not part of the supplementary material~\cite{EOS-DATA-2026-05};
however, the scripts provided in the supplementary material will generate them on demand.
We fit each set with the naive and with the
subtracted detector-level PDF, four fits in all, using random-slice sampling, and stop the nested
sampling at an effective sample size of $2000$. For each parameter,
we determine the posterior probability $P_<$ below the generating value and convert it to
$z = \sqrt{2}\,\mathrm{erf}^{-1}(2P_< - 1)$, i.e., to the quantile of the standard normal distribution
at $P_<$. For a correct posterior, $z$ follows a standard normal distribution.

We show the posteriors for the Crystal Ball resolution in \cref{fig:implementation:B2-corner}; those for
the Gaussian resolution are part of the supplementary material~\cite{EOS-DATA-2026-05}. In all
four fits, $S_0$, $|T_0|/S_0$, and $\arg T_0$ are strongly degenerate along a curved band, since the
angular distribution constrains them mainly through the two combinations $F_H$ and $A_\text{FB}$, such that their
marginal posteriors are wide and far from Gaussian.\footnote{%
    The non-Gaussianity is a purposeful consequence of the choice of benchmark problem and not the subtraction approach.
    The physics application that inspires this benchmark faces the same obstacle.
}
With the subtracted detector-level PDF, the generating
values of all parameters lie within $|z| \leq 0.8$ for both resolutions. For the Gaussian
resolution, $|R_1|$ is only weakly constrained, since little of the $J/\psi$-like peak reaches the window,
and the naive fit reproduces the medians of the subtracted one to within $0.5$ half-widths of
their $68\%$ intervals. For the Crystal Ball resolution, the subtracted fit yields
$|R_1| = 0.763$, with the $68\%$ interval $[0.731, 0.791]$, in agreement with the
generating value $0.784$. The naive fit instead overestimates $|R_1|$ by about $50\%$, the upper bound of its prior:
$92\%$ of its posterior lies within $5\%$ of the prior range from that bound. Its $|z| \simeq 6.5$ is therefore only a lower bound on the
discrepancy. The naive fit thus compensates for the probability that the tails of the resolution carry
from the unresolved $J/\psi$-like peak into the window, which the naive convolution underestimates.

\begin{figure}[!t]
    \centering
    \includegraphics[width=.8\textwidth]{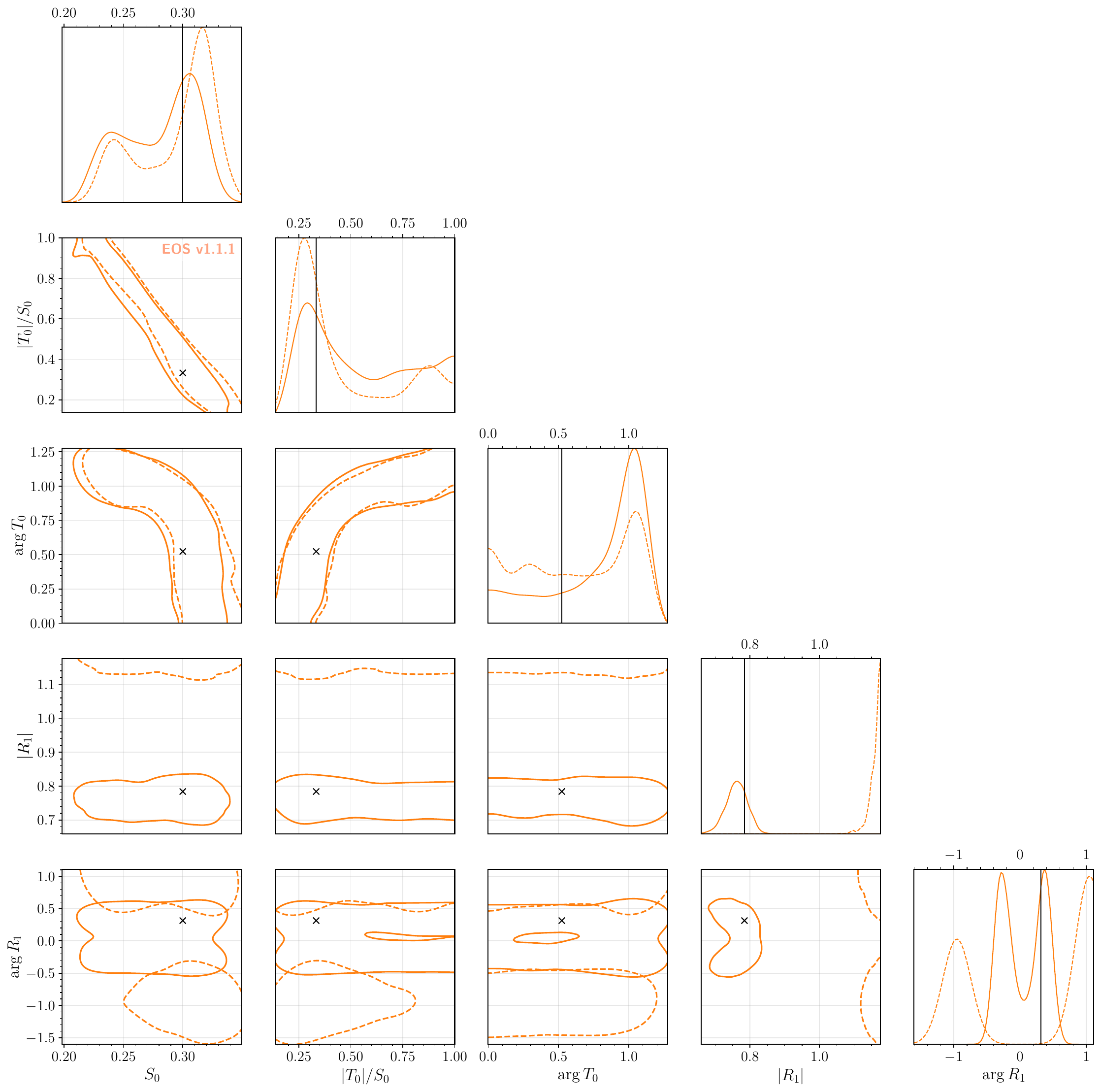}
    \caption{Marginal posteriors of the fits of \cref{benchmark:B2} to the large pseudo-data set with the
    Crystal Ball resolution, with the subtracted (solid) and the naive (dashed) detector-level PDF. The
    two-dimensional panels show the $95\%$ contours. The generating values are marked in black.}
    \label{fig:implementation:B2-corner}
\end{figure}

\noindent
\textbf{Simulation-based calibration.}~%
We validate unbinned fits to \cref{benchmark:B2} within Bayesian statistics, by means of simulation-based
calibration~\cite{Talts:2018zdk}.
We carry out the calibration with the naive and with the subtracted detector-level PDF for the Crystal Ball resolution\footnote{%
    We do not carry out any calibration for the Gaussian resolution, since both approaches to compute its detector-level PDFs
    agree with each other within the $q^2$ window.
}, in campaigns of $500$ replicates each.
For each replicate $n$, we draw a parameter point $\vartheta^{(n)}$
from the prior of \cref{eq:benchmarks:B2:priors}, generate $2.8\cdot 10^6$ truth-level events over the full range, i.e., about as many as are
needed to match the number of LHCb candidates within the $q^2$ window~\cite{LHCb:2026suh} at the
generating point (see \cref{benchmark:B2}), smear the events with the detector resolution, and keep those within the
$q^2$ window. This amounts to a median of $1.08\cdot 10^4$ detector-level events within the window, depending on $\vartheta^{(n)}$
and the choice of resolution. We sample the
posterior given this pseudo-data set by dynamic nested sampling with \texttt{dynesty}~\cite{Speagle:2019ivv};
details on the sampling settings are available from the supplementary material~\cite{EOS-DATA-2026-05}.
For each free parameter, we then determine the rank of $\vartheta^{(n)}$ among $99$
equally weighted posterior draws, and histogram the ranks in $10$ bins. The naive and the subtracted detector-level PDFs are fitted to
identical pseudo-data sets. Since the replicates are drawn from the same prior that enters the posterior, these
ranks are uniformly distributed if the posterior is computed correctly~\cite{Talts:2018zdk}, independently of the
size of the pseudo-data sets. A systematically wrong likelihood, such as one that results from an
inaccurate convolution with the resolution, distorts the distribution of the ranks: a bias skews it,
while posteriors that are too narrow or too wide make it U-shaped or peaked, respectively.
We evaluate the ranks without folding any phase.

\begin{figure}[t]
    \centering
    \includegraphics[width=.8\textwidth]{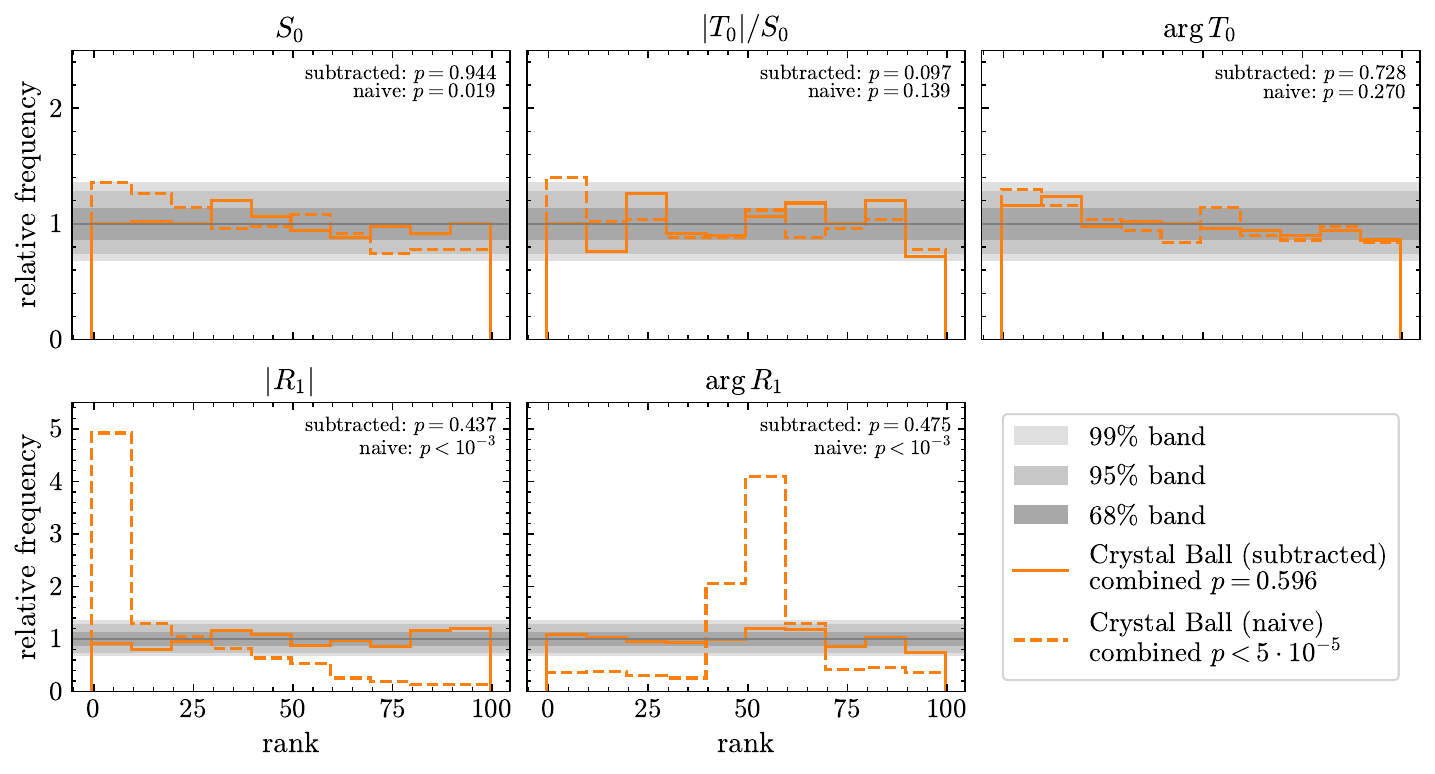}
    \caption{Rank histograms of the simulation-based calibration of \cref{benchmark:B2} with the Crystal Ball
    resolution, for the subtracted (solid) and the naive (dashed) detector-level PDF, with the $68\%$, $95\%$, and
    $99\%$ bands of a flat distribution. The $p$-values refer to the $\chi^2$ test of each histogram.}
    \label{fig:implementation:B2-sbc-ranks}
\end{figure}

We perform two $\chi^2$-based tests for uniformity of the ranks and show their results alongside the underlying histograms in \cref{fig:implementation:B2-sbc-ranks}.
In the first test, we compare each histogram against a flat distribution. For the second test, we combine the histograms of all five parameters by summing their
$\chi^2$ values. Since the ranks of one replicate stem from the same posterior, they are correlated, e.g., those
of $S_0$ and $|T_0|/S_0$ are correlated through the degeneracy of their posteriors discussed above.
To account for these correlations, we obtain the distribution of the total $\chi^2$ for flat ranks from
$2\cdot 10^4$ simulated campaigns, whose ranks carry the correlations measured among the replicates through a Gaussian copula.
For the subtracted detector-level PDF, the calibration succeeds: the rank histograms of all parameters are compatible with flat distributions, with
$p$-values between $0.10$ and $0.94$, and the combined $p$-value is $0.60$.
For the naive detector-level PDF, the calibration fails: the combined $p$-value is below
$5\cdot 10^{-5}$, and $17$ of the $50$ bins lie outside the $99\%$ band of a flat distribution.
The ranks of $|R_1|$ accumulate in the lowest bins, i.e., the naive fits overestimate $|R_1|$, which is compatible
with our findings from the fit to the large pseudo-data set. The ranks of $\arg R_1$ accumulate in the middle bins, i.e., its posteriors place
their mass on both sides of the generating value: as in the fit to the large pseudo-data set, the naive detector-level PDF moves the two modes of
$\arg R_1$, related by the approximate symmetry $\arg R_1 \to -\arg R_1$, apart to about $\pm 1$, such that the
generating value lies between them; see \cref{fig:implementation:B2-corner}. The ranks of $S_0$ are tilted towards
low values ($p = 0.019$), a weaker overestimate. Those of $|T_0|/S_0$ and $\arg T_0$ remain
compatible with flat distributions.

\noindent
\textbf{Computational cost.}~%
One evaluation of the log-likelihood with the subtracted detector-level PDF takes about $0.8$\,s on a
single core, and $0.2$\,s with eight threads. It is dominated by the evaluation of \Ptruth on the grid,
$0.8$\,s, while the sum over the $10^4$ events in the window takes $5$\,ms. A replicate requires
a median of $1.9\cdot 10^4$ evaluations. At the spacing in $m_{\mu\mu}$ at which the naive detector-level
PDF of \cref{benchmark:B1} reaches the accuracy of the subtracted one at the $J/\psi$-like peak, the naive grid would comprise $179200 \times 512 \approx 9.2\cdot 10^7$ points in $(m_{\mu\mu}, \cos\theta_\ell)$. Our implementation holds one observable object per grid point for the truth-level PDF and for the sampled resolution, about $20$\,KiB per point within a fit, and would thus require about $1.7$\,TiB of memory; the numerical arrays of one copy alone would need about $10$\,GiB. Independently of the memory, we estimate one evaluation at about $17$\,s with eight threads, about $90$ times that of the subtracted
detector-level PDF, i.e., about four days per replicate.
For comparison, with the realistic \Ptruth introduced in \cref{sec:implementation:B1}, which takes
$16\,\mu\mathrm{s}$ per point, the evaluation of the subtracted detector-level PDF on the grid of
\cref{benchmark:B2} would take about $17$\,s on a single core and about $4$\,s with eight threads.

\section{Conclusion}
\label{sec:conclusion}

We have presented a method for the convolution of truth-level PDFs that contain narrow poles with
a detector resolution, as required for unbinned likelihood fits to detector-level data. For each pole,
it subtracts a single simple-pole term that collects the squared pole term and the singular part of
its interference, convolves this term through the Cauchy transform of the resolution, and convolves
the smooth remainder numerically on a grid whose spacing is set by the resolution and the nonresonant
amplitude alone. The method extends to multidimensional PDFs, to coefficients obtained from the PDF
on the real axis, to poles in $q^2$ with a resolution in $m_{\mu\mu}$, and to piecewise resolutions.

To test the accuracy of the method, we perform a one-dimensional benchmark that resembles the narrow resonances in the $B^+\to K^+\mu^+\mu^-$
spectrum. The subtraction reduces the relative error at the $J/\psi$-like peak by at least five
orders of magnitude compared to the naive convolution on the same grid, which deviates by up to
$74\%$. For a Gaussian resolution, the subtracted result agrees with the exact one to machine
precision, since the constant remainder of this benchmark is convolved without error; for the
Crystal Ball resolution, its error does not exceed $6.4\cdot 10^{-4}$ away from the edges of the phase
space, limited by the truncation of the expansion of its Cauchy transform. Normalised by its own
integral, the naive result is wrong by a factor of about $3$, and with the Crystal Ball resolution by
up to $43\%$ even outside the standard charmonium windows. A finer grid reaches the accuracy of our
approach at the $J/\psi$-like peak only at about $30$ times the computational cost for $10^4$ events; over
the full range, its error saturates at about $10^{-3}$ unless the padding is enlarged.

To test the performance of the method and to validate it within an unbinned likelihood fit, we perform a two-dimensional benchmark,
again inspired by the $B^+\to K^+\mu^+\mu^-$ spectrum.
Using our method, a grid spacing of half the resolution suffices, since the detector-level PDF agrees with that on a grid twice as fine to within $2\cdot 10^{-5}$ for
$|\cos\theta_\ell| < 0.9$, and to within $5\cdot 10^{-4}$ close to the edges at $\cos\theta_\ell = \pm 1$,
whereas the naive convolution at comparable accuracy would require a grid of about $9.2\cdot 10^7$ points. In fits within a $q^2$ window that
excludes the resonances, the naive convolution with the Crystal Ball resolution still fails, since the
tails of the resolution carry the unresolved $J/\psi$-like peak into the window; in a fit to ten times
the LHCb Run~1+2 sample, it overestimates the residue of the $J/\psi$-like pole by about $50\%$.
Our approach yields posteriors that pass simulation-based calibration at grid sizes governed by the detector resolution,
whereas the naive approach does not.

Our approach is implemented in \EOS version 1.1.1 and ready for application to the unbinned likelihood
of the $B^+\to K^+\mu^+\mu^-$ candidates published by the LHCb collaboration~\cite{LHCb:2026suh},
including its piecewise resolution model. It applies more generally to any unbinned fit of a narrow
resonance that interferes with a nonresonant amplitude, e.g., the Higgs boson in the diphoton spectrum,
whose width of about $4\MeV$ lies far below the experimental resolution and whose interference with
the continuum shifts its apparent mass~\cite{Dixon:2003yb,Martin:2012xc}.

\acknowledgments

We thank Ulrik Egede for discussions about the LHCb dataset,
and Nico Gubernari \& Konstantinos Petridis for discussions and helpful comments on the manuscript.
We further thank the Mainz Institute for Theoretical Physics (MITP) within the PRISMA+ Cluster of Excellence (Johannes Gutenberg University Mainz)
for its hospitality and support during the early stages of this work.
The work of DvD is supported by the UK Science and Technology Facilities Council (STFC) under
grants ST/V003941/1, ST/T001011/1, and ST/X003167/1.
MV has been funded by the Deutsche Forschungsgemeinschaft (DFG, German Research Foundation) --- Emmy Noether Grant No. 558599025.

\appendix
\crefalias{section}{appendix}

\section{Benchmark Problems}
\label{app:benchmarks}

Here, we collect the definitions and further details for our two benchmark problems.

\begin{benchmark}
\label{benchmark:B1}
This benchmark mimics some of the properties of the dimuon mass spectrum of $B^+\to K^+\mu^+\mu^-$ decays in $\xt = m_{\mu\mu}$
(measured in units of \GeV). The nonresonant amplitude is constant, two poles mimic the $J/\psi$ and $\psi(2S)$ resonances, and
the phase space restricts the support of the PDF to the interval $0 \leq \xt \leq W$ with $W = 4.6$.
All components are simple enough that we can compute the convolution with the detector resolutions exactly, up to numerical precision,
and compare these reference results with both the naive and the subtracted DFT-based convolution.

The PDF is constructed following \cref{eq:framework:notation:PDF-decomp} with
\begin{equation}
\label{eq:benchmarks:B1:amplitude}
    \amp(\xt) = 1 + \sum_{j = 1}^{2} \frac{R_j}{\xt - z_j}\,,
    \qquad
    z_j = m_j - \tfrac{i}{2}\Gamma_j \equiv m_j - i\gamma_j\,,
    \qquad
    \Phi(\xt) = \theta(\xt)\,\theta(W - \xt)\,.
\end{equation}
The masses $m_j$ and full widths $\Gamma_j$ approximate those of the two charmonium states.
Since the poles lie far from the edges of the phase space,
$\Phi$ is constant, and thus analytic, in their vicinity, as required in \cref{sec:framework:notation}. We choose the residues such that the ratios
\begin{equation}
\label{eq:benchmarks:B1:ratios}
    r_j \equiv \frac{\int_0^W \mathrm{d}\xt\, |R_j|^2/|\xt - z_j|^2}{\int_0^W \mathrm{d}\xt\, 1}
        \simeq \frac{\pi\,|R_j|^2}{\gamma_j\,W}
\end{equation}
reproduce the ratios of the measured branching fractions,
\begin{equation}
\label{eq:benchmarks:B1:BR}
    r_j = \frac{\mathcal{B}(B^+\to K^+ V_j)\,\mathcal{B}(V_j \to \mu^+\mu^-)}{\mathcal{B}(B^+\to K^+\mu^+\mu^-)}\,,
    \qquad V_1 = J/\psi\,,\quad V_2 = \psi(2S)\,,
\end{equation}
where we neglect the interference terms, which change the integrals only at the level of
$2\gamma_j/|R_j| \lesssim 10^{-2}$. The phases of the residues and the resulting magnitudes are listed in
\cref{tab:benchmarks:B1}.
\begin{table}[t]
\setlength{\tabcolsep}{10pt}
\centering
\begin{tabular}{lllccc}
    \toprule
    Benchmark & quantity                  & unit              & smooth part       & $J/\psi$ ($j = 1$)    & $\psi(2S)$ ($j = 2$)  \\
    \midrule
    \multirow{3}{*}{B1 and B2}
    & $m_j$                               & $\mathrm{GeV}$    &                   & $3.1$                 & $3.7$                 \\
    & $\Gamma_j$                          & $\mathrm{keV}$    &                   & $100$                 & $300$                 \\
    & $r_j$                               & ---               &                   & $134$                 & $11$                  \\
    \midrule
    \multirow{3}{*}{B1}
    & nonresonant amplitude               & ---               & $1$               &                       &                       \\
    & $|R_j|$                             & $\mathrm{MeV}$    &                   & $99.0$                & $49.2$                \\
    & $\arg R_j$                          & ---               &                   & $\pi/10$              & $-4\pi/5$             \\
    \midrule
    \multirow{6}{*}{B2}
    & $V_0$                               & ---               & $1$               &                       &                       \\
    & $S_0$                               & ---               & $0.3^\ast$        &                       &                       \\
    & $|T_0|/S_0$                         & ---               & $1/3^\ast$        &                       &                       \\
    & $\arg T_0$                          & ---               & $\pi/6^\ast$      &                       &                       \\
    & $|R_j|$                             & $\mathrm{GeV}^2$  &                   & $0.784^\ast$          & $0.645$               \\
    & $\arg R_j$                          & ---               &                   & $\pi/10^\ast$         & $-4\pi/5$             \\
    \bottomrule
\end{tabular}
\caption{Parameters of \cref{benchmark:B1,benchmark:B2}. The resonance parameters $m_j$ and $\Gamma_j$ are shared by both
benchmarks and fixed in \cref{benchmark:B2}. The values of $m_j$, $\Gamma_j$, and $r_j$ do not
reflect the exact values compiled in the PDG review~\cite{ParticleDataGroup:2026mpi}, but provide a reasonable approximation of those
found in nature. The ratios $r_j$ are those of \cref{benchmark:B1}; \cref{benchmark:B2} reproduces them to
within $1\%$. Asterisks mark the free parameters of \cref{benchmark:B2}; their values are used to generate the pseudo-data set of
\cref{sec:implementation:B2}.}
\label{tab:benchmarks:B1}
\label{tab:benchmarks:B2}
\end{table}

We confront this benchmark with two resolutions that share a common width
parameter $\sigma = 5\MeV$, a typical value within the range of $2$ to $7\MeV$ quoted in
Ref.~\cite{LHCb:2026suh}:
\begin{itemize}
    \item the Gaussian $\resR[G]$, with $\sigma$ as the sole parameter; and
    \item the two-sided Crystal Ball function $\resR[CB]$ as defined in
    \cref{app:crystal-ball}, with parameters $\alpha_\text{L} = 1.5$,
    $n_\text{L} = 3$, $\alpha_\text{R} = 2$, and $n_\text{R} = 5$.
\end{itemize}
\end{benchmark}

\begin{benchmark}
\label{benchmark:B2}
This benchmark mimics the two-dimensional distribution of $B^+\to K^+\mu^+\mu^-$ decays in $q^2 = m_{\mu\mu}^2$ and the
lepton helicity angle $\cos\theta_\ell$~\cite{Bobeth:2007dw}, without reproducing it in detail. The truth-level PDF reads
\begin{equation}
\label{eq:benchmarks:B2:PDF}
    P_\text{t}(q^2, \cos\theta_\ell) = \frac{\Phi(q^2)}{\mathcal{N}} \left[
        \frac{3}{4}\, |a_V(q^2)|^2 \big(1 - \cos^2\theta_\ell\big)
        + \frac{1}{2} \big|a_S(q^2) + a_T(q^2) \cos\theta_\ell\big|^2 \right]\,.
\end{equation}
The three amplitudes mimic the couplings to the vector, scalar, and tensor lepton currents; the latter
two enter through the combination $a_S + a_T \cos\theta_\ell$, in analogy to the combination
$F_S + \cos\theta\, F_T$ of Ref.~\cite{Bobeth:2007dw}. Equivalently,
\begin{equation}
\label{eq:benchmarks:B2:PDF-observables}
    P_\text{t}(q^2, \cos\theta_\ell) = \frac{\Phi(q^2)\,\rho(q^2)}{\mathcal{N}} \left[
        \frac{3}{4}\big(1 - F_H(q^2)\big)\big(1 - \cos^2\theta_\ell\big)
        + \frac{1}{2} F_H(q^2)
        + A_\text{FB}(q^2) \cos\theta_\ell \right]\,,
\end{equation}
where the rate density $\rho$, with $\mathrm{d}\Gamma/\mathrm{d}q^2 \propto \Phi\,\rho$, the flat term $F_H$, and the
forward-backward asymmetry $A_\text{FB}$ read
\begin{equation}
\label{eq:benchmarks:B2:observables}
    \rho = |a_V|^2 + |a_S|^2 + \frac{1}{3}|a_T|^2\,,
    \qquad
    F_H = \frac{|a_S|^2 + |a_T|^2}{\rho}\,,
    \qquad
    A_\text{FB} = \frac{\Re\left[a_S\, a_T^*\right]}{\rho}\,.
\end{equation}
Each
nonresonant amplitude carries a pole at the mass of the lowest $\bar{b}s$ meson with the
appropriate quantum numbers, as is typical of the respective hadronic form factors:
\begin{equation}
\label{eq:benchmarks:B2:amplitudes}
    a_V(q^2) = \frac{V_0}{1 - q^2/m_{B_s^*}^2} + \sum_{j = 1}^{2} \frac{R_j}{q^2 - s_j}\,,
    \qquad
    a_T(q^2) = \frac{T_0}{1 - q^2/m_{B_s^*}^2}\,,
    \qquad
    a_S(q^2) = \frac{S_0}{1 - q^2/m_{B_{s0}}^2}\,,
\end{equation}
with $s_j = m_j^2 - i\, m_j \Gamma_j$. The charmonium-like resonances couple to the leptons only
through the vector current, as appropriate for their electromagnetic decays to $\mu^+\mu^-$.
They therefore enter $\rho$, but not the numerator of $A_\text{FB}$, which is consequently strongly
suppressed on the resonances. Since $a_V$ does not interfere with $a_S$ and $a_T$, the overall phases of $a_V$ and of the pair
$(a_S, a_T)$ are separately unphysical; we choose $V_0$ and $S_0$ real and positive. The PDF then depends on $\arg T_0$ only through $\cos\arg T_0$. The truth-level PDF is non-negative by construction, for all values of the
parameters; in particular, $|A_\text{FB}| \leq F_H/2$.
The phase-space factor is that of
$B \to K V(\to \mu^+\mu^-)$,
\begin{equation}
\label{eq:benchmarks:B2:phase-space}
    \Phi(q^2) = \lambda^{3/2}(m_B^2, m_K^2, q^2)\, \beta_\mu(q^2)\,,
    \qquad
    \beta_\mu(q^2) = \sqrt{1 - 4 m_\mu^2/q^2}\,,
\end{equation}
which vanishes at both kinematic endpoints, $q^2 = 4 m_\mu^2$ and $q^2 = (m_B - m_K)^2$, and is
smooth in the vicinity of the resonances. The angular part of \cref{eq:benchmarks:B2:PDF-observables} integrates to one, such that
$\mathcal{N} = \int \mathrm{d}q^2\, \Phi(q^2)\,\rho(q^2)$.

The detector resolution acts on $m_{\mu\mu}$ and on $\cos\theta_\ell$. We therefore work in the
variables $(m_{\mu\mu}, \cos\theta_\ell)$, absorb the Jacobian $2 m_{\mu\mu}$ into $\Phi$, and map
the poles in $q^2$ onto poles in $m_{\mu\mu}$ at $z_j = \sqrt{s_j}$, as described in
\cref{sec:framework:complications}. The coefficients $c_j$ then depend on $\cos\theta_\ell$.
In $m_{\mu\mu}$, we use either the Gaussian or the Crystal Ball resolution of \cref{benchmark:B1}; in
$\cos\theta_\ell$, we use a Gaussian resolution of width $\sigma_{\cos\theta} = 0.02$.

The masses $m_j$ and widths $\Gamma_j$ of the resonances, and the phases of their residues, are shared
with \cref{benchmark:B1}. The magnitudes $|R_j|$ are chosen such that the ratios $r_j$ of
\cref{benchmark:B1} are reproduced approximately, with $r_1 \approx 134.6$ and $r_2 \approx 11.1$, where now each ratio is formed from the integrals of $\Phi\,|R_j|^2/|q^2 - s_j|^2$
and of $\Phi\,\rho$ without the resonances over the full phase space. The masses of the initial-
and final-state particles and of the form-factor poles are fixed to values rounded to three decimals; the
latter correspond to the $B_s^*$ and $B_{s0}(5700)$~\cite{LHCb:2026knl} mesons. The nonresonant
parameters yield $F_H \approx 0.09$ and $A_\text{FB} \approx 0.02$ away
from the resonances. The relevant parameters are listed in \cref{tab:benchmarks:B2}.

We fit \cref{benchmark:B2} only to the events within the $q^2$ window
$1\GeV^2 \leq q^2 \leq 8\GeV^2$, i.e., $1.0\GeV \leq m_{\mu\mu} \leq 2.83\GeV$, with $R_2$ fixed
to its generating value for simplicity. The $q^2$ window is sensitive to the low-$q^2$ tail of the $J/\psi$-like amplitude and ends
at the lower edge of the standard charmonium window around it. Unlike Ref.~\cite{LHCb:2026suh}, the
benchmark uses a single resolution throughout; the $q^2$ window contains the boundary at
$m_{\mu\mu} = 1.8\GeV$ between two of its resolution regions. As priors for the free parameters, we use
uniform distributions for
\begin{align}
\label{eq:benchmarks:B2:priors}
    S_0                             & \in [0.1, 0.6]\,, & |T_0|/S_0 & \in [0, 1]\,,      & \arg T_0 & \in [0, \pi]\,, \nonumber\\
    |R_1|/|R_1|_\text{nominal}      & \in [0.5, 1.5]\,, & \arg R_1  & \in [-\pi, \pi]\,. &          &
\end{align}
The parameter point of \cref{tab:benchmarks:B2} lies inside the prior.
The range of $\arg T_0$ exploits the dependence on $\cos\arg T_0$ only, noted above.
The sign of $\arg R_1$ enters the likelihood only through terms of relative order $m_1 \Gamma_1/|R_1|$,
which are too small to be resolved, and through the interference with the fixed $\psi(2S)$-like pole,
which breaks the symmetry under $\arg R_1 \to -\arg R_1$ only weakly. This mirrors the four-fold
ambiguity of the charmonium phases in Ref.~\cite{LHCb:2026suh}. We keep the full range of $\arg R_1$.

We express the size of the pseudo-data sets in units of the LHCb Run~1+2 sample of
Ref.~\cite{LHCb:2026suh}, which comprises $3\,363\,943$ candidates with
$300 \leq m_{\mu\mu} \leq 4700\MeV$, $8\,088$ of which lie within the $q^2$ window. One unit
corresponds to $2.8\cdot 10^6$ events of \cref{benchmark:B2} generated over the full range,
about $8.0\cdot 10^3$ of which lie within the $q^2$ window for the Gaussian and about
$8.5\cdot 10^3$ for the Crystal Ball resolution. Generating the events over the full range and
keeping those within the window reproduces the leakage of the $J/\psi$-like peak into the window
through the tails of the resolution, which accounts for the larger number for the Crystal Ball
resolution.
\end{benchmark}

\section{Cauchy Transform of the Crystal Ball Function}
\label{app:crystal-ball}

The original Crystal Ball function~\cite{Oreglia:1980cs,Gaiser:1982yw,Skwarnicki:1986xj} combines a Gaussian core with a single power-law tail on its low side; its
two-sided generalisation adds a second, independent power-law tail on the high side. For width $\sigma$, it is defined as
\begin{equation}
\label{eq:app:cb:def}
    \resR[CB](u) = \frac{1}{\mathcal{N}_\text{CB}}
    \begin{cases}
        A_\text{L} \left(B_\text{L} - u/\sigma\right)^{-n_\text{L}}\,, & u/\sigma < -\alpha_\text{L}\,,\\
        e^{-u^2/(2\sigma^2)}\,, & -\alpha_\text{L} \leq u/\sigma \leq \alpha_\text{R}\,,\\
        A_\text{R} \left(B_\text{R} + u/\sigma\right)^{-n_\text{R}}\,, & u/\sigma > \alpha_\text{R}\,,
    \end{cases}
\end{equation}
with $A_s = (n_s/\alpha_s)^{n_s}\, e^{-\alpha_s^2/2}$ and $B_s = n_s/\alpha_s - \alpha_s$ for
the left and right tails, $s = \text{L}, \text{R}$.
The resolution $\resR[CB]$ is continuously differentiable at the transition points.
Its normalisation is
\begin{equation}
    \mathcal{N}_\text{CB} = \sigma \left[ \sqrt{\frac{\pi}{2}} \left(
        \operatorname{erf}\frac{\alpha_\text{L}}{\sqrt{2}} + \operatorname{erf}\frac{\alpha_\text{R}}{\sqrt{2}}
        \right)
        + \sum_{s = \text{L}, \text{R}} \frac{n_s\, e^{-\alpha_s^2/2}}{\alpha_s (n_s - 1)} \right]\,.
\end{equation}

An exact evaluation of $\cauchy{\resR[CB]}$ is possible in principle. The power-law tails yield
elementary Cauchy transforms (partial fractions and logarithms for integer $n_s$, and
hypergeometric functions ${}_2F_1$ otherwise). The truncated Gaussian core, however, requires
a generalisation of the Faddeeva function to an ``incomplete'' Faddeeva function
$\int_{a}^{b}\!\mathrm{d}t\, e^{-t^2/2}/(\zeta - t)$.
We are not aware of an implementation of this function of a quality comparable to that of the
Faddeeva function in the open-source \texttt{libcerf} library~\cite{libcerf}.
We therefore refrain from implementing the exact evaluation of $\cauchy{\resR[CB]}$ and pursue a
simpler approach instead.

We split the Crystal Ball function into the Gaussian $\resR[G]$ of unit norm and of the same width
$\sigma$, and a correction term $\Delta$,%
\footnote{%
    Note that $\Delta$ is not a resolution kernel, since it takes negative values and its integral
    over the real axis vanishes. Hence, we do not use the \resR notation for $\Delta$.
}
\begin{equation}
\label{eq:app:cb:split}
    \resR[CB] = \resR[G] + \Delta\,,
    \qquad
    \Delta(u) =
    \begin{cases}
        \mathcal{N}_\text{CB}^{-1} A_\text{L} \left(B_\text{L} - u/\sigma\right)^{-n_\text{L}} - \resR[G](u)\,, & u/\sigma < -\alpha_\text{L}\,,\\
        \big[\mathcal{N}_\text{CB}^{-1} - (\sqrt{2\pi\sigma^2})^{-1}\big]\, e^{-u^2/(2\sigma^2)}\,, & -\alpha_\text{L} \leq u/\sigma \leq \alpha_\text{R}\,,\\
        \mathcal{N}_\text{CB}^{-1} A_\text{R} \left(B_\text{R} + u/\sigma\right)^{-n_\text{R}} - \resR[G](u)\,, & u/\sigma > \alpha_\text{R}\,.
    \end{cases}
\end{equation}
In the core, $\Delta$ is a rescaled Gaussian, which does not vanish, since the normalisations of
$\resR[CB]$ and $\resR[G]$ differ. The Cauchy transform is linear in the kernel,
\begin{equation}
\label{eq:app:cb:cauchy-split}
    \cauchy{\resR[CB]} = \cauchy{\resR[G]} + \cauchy{\Delta}\,,
\end{equation}
and $\cauchy{\resR[G]}$ is known in closed form in terms of the Faddeeva function. The correction $\Delta$ is continuously differentiable, and its core has
a width of order $\sigma$, much larger than $\gamma$. Since
$\cauchy{\resR}$ is analytic in the upper half-plane with $\mathrm{d}\cauchy{\resR}/\mathrm{d}\zeta
= \cauchy{\resR'}$, and its boundary values on the real axis follow from the Sokhotski--Plemelj
theorem, we expand $\cauchy{\Delta}$ in $\gamma/\sigma$. With $\xd - z_j = y + i\gamma$ and
$y \equiv \xd - x_j$,
\begin{equation}
\label{eq:app:cb:expansion}
    \cauchy{\Delta}(y + i\gamma) = \pi \left[ \hilbert{\Delta}(y) - i\,\Delta(y) \right]
        + i\gamma\,\pi \left[ \hilbert{\Delta'}(y) - i\,\Delta'(y) \right]
        + \mathcal{O}(\gamma^2 \ln \gamma)\,,
\end{equation}
where $\Delta'$ indicates the first derivative of $\Delta$ with respect to its offset variable and
we introduce the Hilbert transform of a function $f$,
\begin{equation}
    \pi\,\hilbert{f}(y) \equiv \operatorname{p.v.}\!\int \mathrm{d}u\, \frac{f(u)}{y - u}\,.
\end{equation}
The functions $\hilbert{\Delta}$ and $\hilbert{\Delta'}$ depend
only on the resolution and not on the parameter point. They are therefore tabulated once per
resolution configuration and interpolated thereafter. Since $\Delta(y)/\Delta(0)$ depends on $y$ only
through $\eta = y/\sigma$, we tabulate $\pi\hilbert{\Delta}$ and $\pi\hilbert{\Delta'}$ as functions of
$\eta$; the tables thus depend only on the tail parameters $\alpha_s$ and $n_s$. They cover $|y|$ up to
twice the extent of the grid, i.e., $|\eta| \lesssim 2\cdot 10^3$ in our benchmarks, on about $3300$ nodes
uniformly spaced in $\operatorname{asinh}(\eta)$, which concentrates them in the core while reaching far
into the tails, and are interpolated by four-point Lagrange interpolation. They must be recomputed
whenever the tail parameters change, which takes about half a second as recorded in\cref{sec:implementation:B1}; a
change of $\sigma$, e.g., of a scale factor calibrated on the $J/\psi$ in the fit itself, requires
extending them only if the range in $\eta$ grows.
We compute $\pi\hilbert{f}(\eta) = \operatorname{p.v.}\!\int \mathrm{d}t\, f(t)/(\eta - t)$ for
$f = \Delta, \Delta'$ at each node by subtracting the singularity: within $|t| \leq T$, we integrate
$[f(t) - f(\eta)]/(\eta - t)$, which is regular, by adaptive quadrature with breakpoints at the junctions, at
$\eta$ and $\eta \pm 1$, and at powers of ten, and add $f(\eta)\ln[(T + \eta)/(T - \eta)]$ analytically; beyond $T$, we
integrate the power-law tails after the substitution $t = \pm T \tau^{-a}$, which maps them onto
$\tau \in (0, 1]$.

For the tail parameters of \cref{benchmark:B1}, we find the following largest relative errors of
$\cauchy{\resR[CB]}$ with respect to adaptive quadrature, over offsets $-10\sigma \leq y \leq 10\sigma$:
\begin{center}
\begin{tabular}{lcc}
    \toprule
    expansion                                               & $\gamma/\sigma = 0.01$    & $\gamma/\sigma = 0.03$    \\
    \midrule
    full kernel, leading order                              & $7.9\cdot 10^{-3}$        & $2.4\cdot 10^{-2}$        \\
    full kernel, including $\mathcal{O}(\gamma)$            & $9.8\cdot 10^{-5}$        & $7.7\cdot 10^{-4}$        \\
    $\Delta$ only, leading order                            & $2.0\cdot 10^{-3}$        & $6.0\cdot 10^{-3}$        \\
    $\Delta$ only, including $\mathcal{O}(\gamma)$          & $5.0\cdot 10^{-5}$        & $3.3\cdot 10^{-4}$        \\
    \bottomrule
\end{tabular}
\end{center}
The $\mathcal{O}(\gamma)$ term reduces the error by a factor of about $20$ to $80$; applying the
expansion to $\Delta$ only, which treats the Gaussian core exactly, gains a further factor of two to
four. The largest errors arise for negative offsets, on the side of the heavier tail
($n_\text{L} = 3$); including the $\mathcal{O}(\gamma)$ term, they occur at the junction
$y = -\alpha_\text{L}\sigma$, where the second derivative of the Crystal Ball function jumps. From
$\gamma/\sigma = 0.01$ to $0.03$, the leading-order errors grow by a factor of three, as expected.
With $\Delta$ only and the $\mathcal{O}(\gamma)$ term, the error grows by a factor of $8.7$ away from the
junctions, as expected for a remainder of order $\gamma^2$, and by $6.6$ at the junction, where
$\Delta''$ jumps, such that the remainder scales like $\gamma^2\ln\gamma$, which predicts $6.85$.
For \cref{benchmark:B1}, with $\gamma/\sigma = 0.01$ and $0.03$, the resulting errors of the detector-level
PDF do not exceed $7.4\cdot 10^{-5}$ and $6.4\cdot 10^{-4}$, respectively (\cref{sec:implementation:B1});
we therefore do not pursue the exact evaluation further.
The size of $\Delta$, and with it the accuracy of the truncated expansion and the padding required, is
set by the power-law tails of the resolution. The low-side tail stems mostly from final-state
radiation and mismeasured muons, and can be reduced by selections on reconstructed quantities, e.g., a
tighter window in the reconstructed $B$ mass, at the price of re-deriving the efficiency and the
resolution from simulation.

\bibliography{references}

\end{document}